\documentclass[journal]{IEEEtran}
\IEEEoverridecommandlockouts
\usepackage{cite}
\usepackage{setspace}
\renewcommand{\baselinestretch}{1} \normalsize
\usepackage{amsmath,amssymb,amsfonts}
\usepackage{algorithmic}
\usepackage{graphicx}
\usepackage{textcomp}
\usepackage{xcolor}
\usepackage{url}
\usepackage{soul}
\usepackage{ulem}

\def\BibTeX{{\rm B\kern-.05em{\sc i\kern-.025em b}\kern-.08em
    T\kern-.1667em\lower.7ex\hbox{E}\kern-.125emX}}
\begin{document}

\title{\huge{When Web 3.0 Meets Reality: A Hyperdimensional Fractal Polytope P2P Ecosystems}
%A Decent Controller for Web3.0 native infrastructure
%When Web3 meets Reality: Hyperdimensional Fractal P2P Ecosystems
%{\footnotesize \textsuperscript{*}Note: Sub-titles are not captured in Xplore and should not be used}
% \thanks{}
}

% \author{Author1
% 	\thanks{Corresponding authors: x}
% 	\thanks{
% 		xxx
		
% 	}}
 
\author{Hao Xu, Yunqing Sun, Xiaoshuai Zhang, Erwu Liu and Chih-Lin I~\IEEEmembership{Fellow,~IEEE}
	% \thanks{Corresponding authors: Yunqing Sun and Xiaoshuai Zhang}
	\thanks{
		 Hao Xu is with Huawei Technologies (UK), Cambridge, CB4 0WG, UK, E-mail: hao.xu@ieee.org. Yunqing Sun is with Department of Computer Science, McCormick School of Engineering and Applied Science, Northwestern University, Evanston, IL, US, E-mail: yunqing.sun@northwestern.edu. Xiaoshuai Zhang is with University of Glasgow, Glasgow, G12 8QQ, UK, E-mail: Xiaoshuai.Zhang@glasgow.ac.uk. Erwu Liu is with College of Electronics and Information Engineering, Tongji University, Shanghai, China; E-mail: erwuliu@tongji.edu.cn. Chih-Lin I is with China Mobile Research Institute, Beijing, China; E-mail: icl@chinamobile.com.
		
	}}
\maketitle

\begin{abstract}
Web 3.0 opens the world of new existence of the crypto-network-entity, which is independently defined by the public key pairs for entities and the connection to the Web 3.0 cyberspace. In this paper, we first discover a spacetime coordinate system based on fractal polytope in any dimensions with discrete time offered by blockchain and consensus. Second, the novel network entities and functions are defined to make use of hyperdimensional deterministic switching and routing protocols and blockchain-enabled mutual authentication. In addition to spacetime network architecture, we also define a multi-tier identity scheme which extends the native Web 3.0 crypto-network-entity to outer cyber and physical world, offering legal-compliant anonymity and linkability to all derived identifiers of entities. In this way, we unify the holistic Web 3.0 network based on persistent spacetime and its entity extension to our cyber and physical world.
\end{abstract}

\begin{IEEEkeywords}
Web 3.0, fractal network, decentralized infrastructure and identity, blockchain, polytope
\end{IEEEkeywords}
\section{Introduction}
Web 3.0 represents the next stage in the development of the World Wide Web by 
% which aims to provide users with equal access to applications, services and multimedia content through decentralized network interfaces. 
encouraging peer-to-peer networks among all users who enjoy equal ownership of their digital assets without relying on centralized servers or third-party intermediaries.
By leveraging the Web 3.0 ready network, entities are both reached locally and globally in a unified access scheme. However, existing computer networking protocols, e.g., IPv4 and IPv6, treat the address with local and global routing based on routing tables within the domain. On the other hand, the current network cannot resolve addresses and identifiers without the records from name servers, which are eventually aggregated as the root Domain Name Servers (DNS) held by central internet agencies and hinders decentralization in the global scale \cite{Lin2021}. With the existing infrastructure, peer-to-peer networks are easily suffocated by tycoons and experiencing interruptions of services, as P2P networks are merely extensions of physical networks, where all devices are connected to central network infrastructures.  

Distributed Ledger Technology (DLT) or Blockchain is considered as the backbone of Web 3.0, which enables information distribution, and protects data integrity and credibility while maintaining consistency using consensus mechanisms \cite{Li2021}. And when it comes to the existence of objects and value in Web 3.0 due to anonymity and its independence from physical networks, there is always a debate about the metaphysical significance of everything on the blockchain and Web 3.0. People often wonder if their assets are real in Web 3.0, or if their identities are indiscernible in this digital realm. Can our identities truly be located in cyberspace with precise spacetime, even though they may not have any physical presence?

\subsection{Motivations of Web 3.0 network and identity management}
Web 3.0 offers a viewpoint of decentralization for reconstructing WWW and the internet services in controversial to big giants and centralized infrastructures with the power of blockchain \cite{bambacht2022web3}.
Blockchain has only been considered as an one-dimensional data structure and has a little reflection of the actual/logical world. In fact, linear and repetitive data blocks resemble the idea of discrete time in Web 3.0.
As the pillar of decentralized ecosystems, the blockchain is not only a chained undeniable crypto vessel but a spatio-temporal description of everything connected and mapped into the topology. 

%$%% $  In this paper, we propose the fundamental spacetime coordination with the skeleton of network built on blockchain which combines a novel network-spatial blockchain in hyper-dimensional manner with the state description of whole network and their identities mapped to the spatial ontological entries. Every object is geometrically mapped to the spatial description of logical space which is agnostic to the real world, defined only by the topological relations and the identity and timestamped by blockchain.  
%区块链作为Web3.0和元宇宙的基础设施，一直被视为一维数据结构，对实际/逻辑世界的反映很少，但数据呈线性和重复增长。区块链不仅是一个不可否认的加密容器，而且是Web3.0中连接一切的空间描述。在本文中，我们使用区块链构建了网络的骨架，将一种新颖的网络空间区块链与整个网络的状态描述相结合，并将它们的身份映射到空间本体条目中。每个对象都被几何地映射到逻辑空间的空间描述中，该描述对真实世界是不可知的，仅由拓扑关系和身份定义。
 
%% In current, the dramatic advance of Web 3.0 has arisen interest from different stakeholders. Meanwhile, numerous Web 3.0 and metaverse applications are blooming based on diverse blockchain/DLT platforms such as decentralized finance (DeFi) and game finance (GameFi) and permanent storage. However, Web 3.0 is still in its infancy since a global Web 3.0 to overturn Web 1.0/2.0 is facing many challenges, especially for network scalability and identity management.

%With a great boost on security, privacy and cyber sovereignty ({\color{black}cyber sovereignty refers to the cyber boundary established by a country or region for exercising national control and implementing specific legislation}) of user data, the challenges faced in achieving Web3.0 and opportunities are significant. 

\subsubsection{Web 3.0 network}
\begin{figure*}[htbp]
    \centering
    \includegraphics[width =1.05\textwidth]{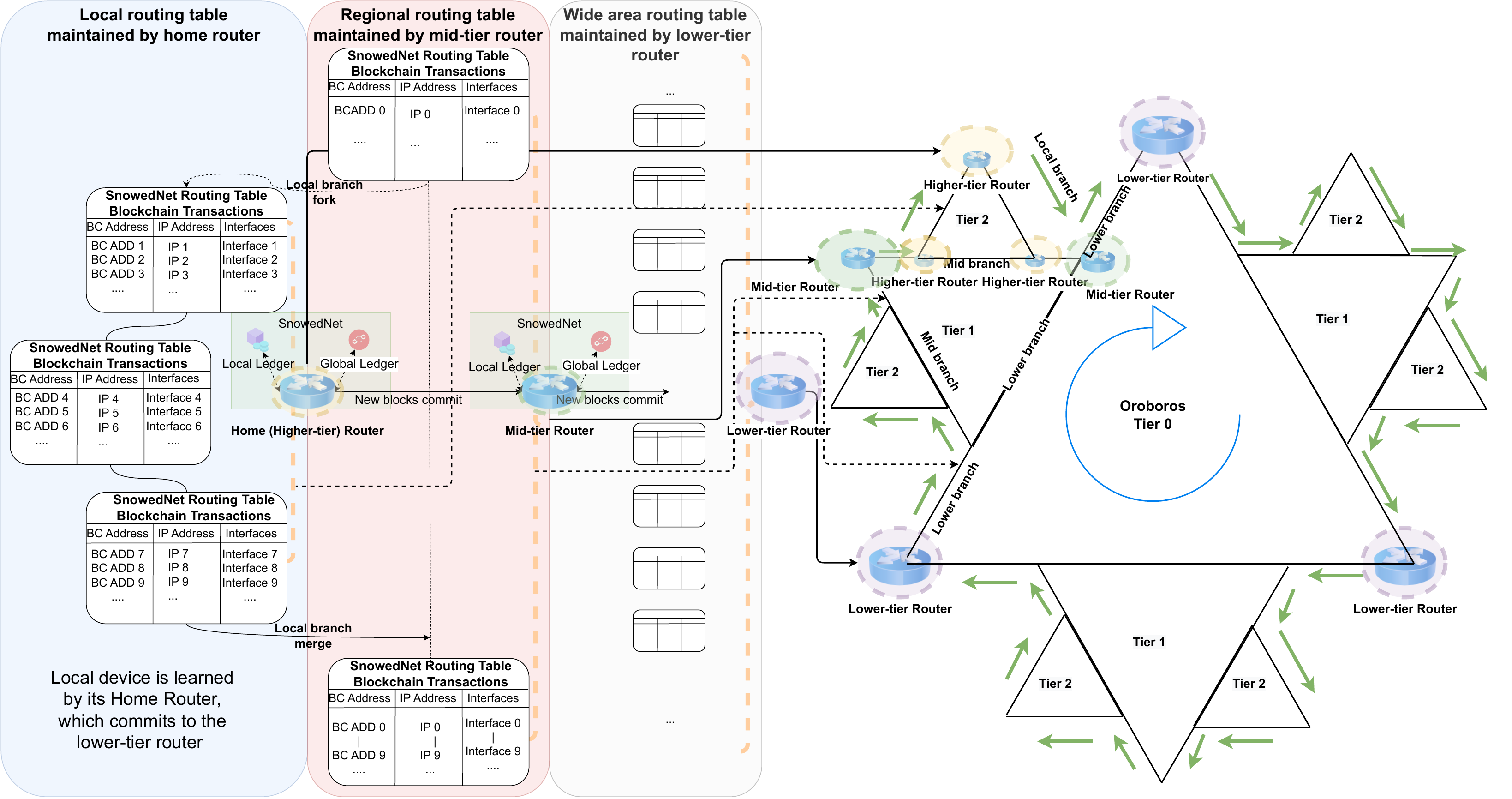}
    \caption{Architecture of the hyperdimensional fractal network for Web 3.0 ($D = 2$).}
    \label{fig:hyper}
    %\vspace{-5px}
\end{figure*}

%for reality mapping: space and time
Based on persistent decentralized medium and data, the Web 3.0 network is not only the extension of reality, but a new ``reality'' is happening within. However, to define an existence, the network needs to provide the persistent and independent coordinates of the existent, which are inherently true when they are connected in any space where the coordinate of the existent can be twofold: the space and the time. Thanks to blockchain-backed network, the network state database, e.g., connectivity, temporal topology and aliveness status, are subject to continuous refresh based on concurrent connections at any given time. An example of network identity mapping into a network address with blockchain timestamp can be found in Fig. \ref{fig:hyper}, where the network is defined as a regular triangle with Koch fractal, more iterations are illustrated in Fig.~\ref{fig:dimension}. 
Compared with the current centralized network paradigm, such a fractal shaped architecture is native globally peer-to-peer and deterministic with shortest paths to better facilitate the future encrypted Web 3.0 network.
%for scalability and latency

Furthermore, the issue of network scalability needs to be addressed to construct a truly global and interplanetary Web 3.0 network. 
Unlike existing networks, the global Web 3.0 network will be a massive entity involving countless users and stakeholders, offering a vast array of services and applications, and processing enormous transactions. This amplifies the need for high network throughput and highlights the negative impact of network latency. Novel networks are yet to concern the consistent global state with high regional throughput, as seen in Fig. \ref{fig:hyper}, where the network shards are committed into the upper network nodes. 

% Furthermore, DLT applied in Web 3.0 requires consensus mechanisms for accepting or declining new committed transactions and synchronizing ledgers, in which consensus latency can also be affected by innumerable nodes within the huge Web 3.0 network. Therefore, a suitable network architecture to Web 3.0 should be sculptured to endorse the core value ``read-write-own" of Web 3.0. 

%however, if the access to the space of Web3.0 is denied, ownership means little or nothing to the owner who is blocked from accessing the WWW. Meanwhile, the value of privacy offered by Web3.0 becomes void if the user can be tracked at the beginning and the end of Internet access. It is necessary to ensure the user will never be unplugged from the network or illegally tracked due to centralization causes. Most importantly, Web3.0 shall secure itself from running the whole network on the infrastructure offered by centralized resource controllers.

\subsubsection{Identity management}
%for mapping the existent in a continous mannar
When it comes to existence, identity matters. The whole network needs to recognize a user's identity and its derivatives in a consistent manner i.e., no matter how the identifiers are processed in cryptography, they can always be linked back to a real identity and the existent in the network. Ghost identities, which are pure cryptographic products not linked to any real entities, shall not be considered a real identity, as there is no spatio-temporal existence of them.

% for identity decentralization and anonymization
Users' identities are managed and verified by third-party applications and service providers i.e., they possess the identity information of their clients. Compared with Web 1.0/2.0, identity management in Web 3.0 needs to be more universal because no centralized entity can harness users' identities \cite{Xu2023}. Furthermore, since all users are equivalent and their identities are kept anonymous in the decentralized Web 3.0 network, a real user identity should only be possessed and controlled by the user itself and the sensitive information used in identity verification needs to be regulated by the user.

% for identity juristiction
% Web 3.0 is the next version of the World Wide Web. It was created to be the successor to Web 2.0. 
Web 3.0 seeks to eliminate privacy and data governance issues in today's Internet. With the features of decentralization and anonymization, users have more control over their personal data and privacy where they own their autonomy to make choices. However, Web 3.0 may bring some new potential challenges in the perspective of legislation such as data jurisdiction and self-sovereign identity, which should be further analyzed in line with the objectives of General Data Protection Regulation (GDPR). In this paper, a novel scheme for the universal identity with their juridical compliance is proposed based on their spatio-temporal existence in Web 3.0.

\subsection{Contribution}
In this paper, we propose a fundamental spacetime coordinate system with the skeleton of a network built on blockchain. This coordinate system combines a novel network-spatial blockchain in a hyperdimensional manner with the state description of the whole network and the identities mapped to the spatial ontological entries \cite{Eriksson2022}. It impacts the existing network by enabling deterministic geometrical mapping of every connected entity to the spatial description of logical space which is agnostic to the real world. Hence, they are defined only by the topological relations and identity while being timestamped by blockchain. 

Our contributions to Web 3.0 lay in two aspects:
\begin{itemize}
    \item A hyperdimensional simplex fractal Web 3.0 network based on blockchain, coined as SnowedNet, offering a discrete spatio-temporal coordinate, which is mapped to the physical world, to connect existent with considering blockchain consensus process and routing strategies;
    
    \item A universal identity management scheme that extends the use of anonymous and consistent Web 3.0 identity in the cyber world and physical world.

    % \item Legal challenges and considerations for Web 3.0 network and identity management.
\end{itemize}

%(a). the Web3.0 network architecture with the detailed description of deController consisting of the overlay and underlay network; (b). the security, privacy and identity in a fully decentralized manner; (c). the operational principles regarding law and governance for Web3.0 infrastructure as shown in Fig. \ref{fig:dimension}.

\section{Hyperdimensional fractal Web 3.0 network}
\begin{figure}[htbp]
    \centering
    \includegraphics[width =0.48\textwidth]{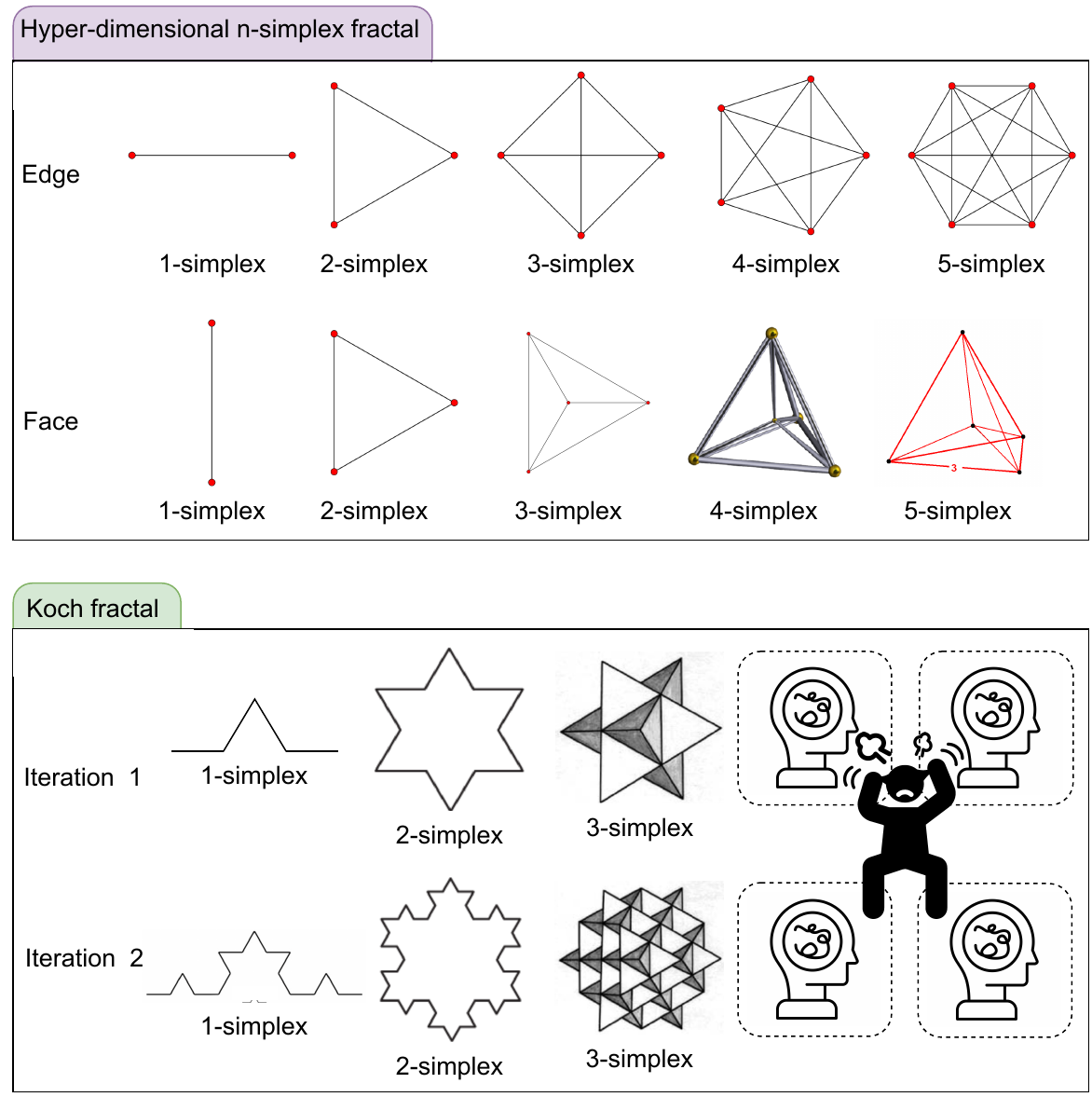} % to be replaced by vector img
    \caption{Hyperdimensional n-simplex fractal edge and face stereophonic projection and Koch fractal iterations (tiers) 1 and 2, note that 4-simplex and 5-simplex cannot be viewed as a whole in 3D, and Koch fractal iterations of n-simplex ($n \geq 4$) are beyond authors' imagination.}
    \label{fig:dimension}
\end{figure}

The first mathematical introduction of fractal shape was first discovered in measuring the length of the coast of Britain. A fractal shape is usually defined as a rough or fragmented geometric shape that can be divided into several parts, and each part is at least approximately a reduced shape of the whole, i.e., it has the property of self-similarity as shown in Fig. \ref{fig:dimension}. A fractal shape is an abstract object in mathematics used to describe things that exist in nature. Inspired by fractal shape and Hausdorff dimension \cite{falconer_1985}, the fractal dimension of the network is proposed to analyze complex networks. As seen in Fig. \ref{fig:hyper}, the network can grow in terms of tiers, which follows the same pattern of the previous step, and the network can have infinite tiers, allowing infinite connections and infinite shards of the network.
% Based on fractal shape, many studies have sculpted the whole Internet as a hyper-dimensional fractal shape.

\subsection{Fractal network overview}
% In the recent advances of blockchain, we have seen applications in Radio Access Network (RAN) \cite{Xu2023decent}, and most importantly, blockchain address-based identity scheme and access control  \cite{Xu2023}, which are important Web 3.0 infrastructures and protocols. 
The blockchain has a timely synchronized mechanism that ensures the information is distributed across nodes with the ability to verify the data integrity efficiently. In the fractal network, the blockchain brings an opportunity to enable the dynamics of label-based routing/switching~\cite{RFC3031}, in the way of updating the routing/switching records in a timely and accountable manner with the reinforced security for transportation and authentication because of encrypted identities. With the given amount of records, if the blockchain records are set to be only kept at the given length and refresh the chain and blocks over time, and old blocks are considered in the last epoch, which would not be synchronized to the networking devices but stored in the archive. New records will replace the old records block by block on a periodical basis.
The behavior of such blockchain is like an \textbf{Ouroboros}, the snake has its tail in its mouth, rolling over and over again. The record which points to an entity in the real world will be set to refresh every time a block is generated, hence becoming a dynamic network client. The blockchain is also considered a global public network, which keeps all records on every participating node. Meanwhile, each fraction or segment of the longest chain can be regarded as a segment of records of what happened on the chain, and verifiable with little effort by distributed nodes regardless of where records are kept on the chain. 

\subsection{Hyperdimensional fractal network architecture for Web 3.0}\label{three-tier}

\subsubsection{Hyperdimensional fractals}
Network is naturally visible as a flat surface with every connection from end to end. However, the use of overlay in the network has created a new mind-blowing topological challenge, with the state-of-the-art software-defined networks (SDN). The SDN controller builds multiple bridges among the peer-to-peer connected entities, for example, VPN and tunneling between servers. The underlying network is responsible for forwarding the network traffic based on distance vectors or link states. Blockchain full nodes are considered as peers in the network, and they are required to be a full replica of each other.

The network is initially established on multi-port switches, by building an equivalent relation between switches with a given number of ports so a node can have direct connections according to quantity of ports, $D$. If all nodes are connected directly, they will form a topological relation where each node has $D$ arms, and each arm has identical jumps among all nodes represented by a unit length in the topology. In the case $D = 1$, we have a segment line with two vertices, known as 1-simplex \cite{McMullen2002}, shown in Fig. \ref{fig:dimension}, and 2-simplex, an equilateral triangle in the case $D = 2$, and a tetrahedron with $D = 3$, 4-simplex, a.k.a., pentachoron, when $D =4$. However, we can have switches with ports as many as 24/48/96 ports, which leads to $D = 24/48/96$ to create a 24/48/96-simplex and forms a near hypersphere in the dimension of 24/48/96. The dimension of the network as a whole is largely decided by the switch with the most ports $D$, and all sub-network can be seen as a $D$ dimensional space reduced to its actual port quantity. It is always true to have a full topological description of the lower dimension structure in a higher dimension. Note that, blockchain consensus and the network throughput might become the bottleneck of achievable dimensions. Consensus, in particular, voting-based consensus, suffers heavy communication overhead and a significant computation complexity challenge. As for the network throughput, the traffic amount on the tier 0 routers is effectively the summit of the whole network. 

Meanwhile, when we attach a sub-network to any nodes, e.g., shards in Fig. \ref{fig:hyper}, it establishes a tiered topology with its hierarchy similar to the parent network. In that case, the network starts growing to higher tiers, no matter which dimension it belongs to. The topology is fractal in both dimensional and geometrical. 

\subsubsection{Design overview - entities and functions}

To sculpt the Web 3.0 network with fractal shapes abstracted from reality, the computer network is mapped into a new network architecture we propose, named SnowedNet with characteristics from blockchain, fractal shapes, security, routing and switching. The entities in SnowedNet are listed as follows.
\begin{itemize}
  \item \textbf{Blockchain address ($\sf{BCADD}$)} is any encoded hashed public key that fits the requirements of blockchain platforms, acting as the encrypted identity, for instance, ERC-20 compatible address or W3C/ISO DID (decentralized identity) compliant if required in the future. The $\sf{BCADD}$ is a prerequisite for anything connected to the network. 
  % \item Interface is of any records directs to a physical or logical outlet for the proclaimed entity.
  
  \item \textbf{SnowedNet nodes}, are an umbrella term for both SnowedNet routers or end devices. The node of SnowedNet is marked with $\sf{BCADD}$. When the node is an end device, it can be routed directly and determined without further lookup of blockchain addresses, as it is a vertices of the topology.  
  
  \item \textbf{SnowedNet records} consist of two basic types of transactions, self-claims and routing requests. They are basic units in both SnowedNet blockchain and SnowedNet routers and switches. For every record, there is an assigned label by the end-point routing node, and the label contains the information of current label-based routing information.
  
  \item \textbf{SnowedNet blocks}, are ordinary blockchain blocks, linked by hash and timestamps.
  
  \item \textbf{SnowedNet shards}, are the composition unit of a given height of segments, the node within the shard keeps the identical records as other peer nodes under the same segments. And the shard represented by interconnected nodes within the same tier is considered as a contention group.
  
  \item \textbf{SnowedNet segments}, the length unit of SnowedNet blockchain, are consisted of blocks with varied lengths. A segment is defined with the number of blocks, known as the capacity of the segment. Moreover, the lower tier always keeps more segments of blocks. 
  
  \item \textbf{SnowedNet addresses}, are hybrid addresses of SnowedNet topology labels and entities' $\sf{BCADD}$ and their other (optional) addresses. A complete SnowedNet address is both globally routable and switchable.

  \item \textbf{Block height} is represented by tier number and the indexes of labels. The tier is counted from 0, as shown in Fig. \ref{fig:hyper}. The block height is limited by a maximum value defined by the tier number and the segment capacity.
  
  \item \textbf{Routing path }is the outcome of the calculated shortest paths between the requested client and the origin node. The path is dynamic and real-time based on the latest routing information shared by its upper-tier routers. The path can always be obtained by the upper-tier router, as it knows the whole composition of current sub-nets with all labels under its hierarchy.  
  
  \item \textbf{SnowedNet epoch}, is a preset time period for a complete refresh in the direction marked by arrows in Fig. \ref{fig:hyper}. SnowedNet refresh makes the original blockchain folded as a coil. All records are reeled with the latest records always on the surface, resembling the blockchain into a block ring. 
 
%  \item \textbf{BeMutual Authentication} is designed in accordance with Blockchain-enabled Network \cite{Xu2021beran}, the authentication happens between two blockchain addresses in order to enable certificate-less zero-trust mutual authentication over blockchain network. The inherit methodology can be perfectly integrated with SnowedNet routing and switching.

\end{itemize}
% The Web3.0 features the owner economy, which boosts the decentralized applications (dApp). The dApp is a smart contract powered autonomous code running on decentralized networks. Once the code is deployed on the blockchain, it becomes a public asset for
% on the blockchain, and can be invoked by 
% any entities within the network. However, the dApp only works as an agent passing on the value between users; it cannot offer demanding services, e.g., video streaming, chatting room or online gaming. To enrich the context of Web3.0 ecosystem, service providers can use dApp to securely provide services to users using encrypted identities and exchange tokens, hence becoming a decentralized service provider. 

\subsubsection{Consensus process}

The consensus for maintaining the blockchain network is proof-of-routes (length). The router with most routes should have a greater chance to mine. Meanwhile, the winning nodes are under continuous testing of the routing (or switching) speed, by evaluating searching algorithms to validate its performance. A threshold can be set up for each router's minimum searching performance in each tier, as the consensus should maintain its minimum commitment to deterministic network and end-to-end latency.

The nodes in vertices or the routing node are set to be re-elected during the mining activities in each tier, and the winner node can replace its lower-tier node in a competitive manner. In addition, the unsuccessful node will be set back to the higher tier.

To obtain the maximum live routes, the following steps of sending $\sf heartbeat$ packet can be used:

\begin{enumerate}
    \item The node unicasts to all neighboring nodes, as one of the requirements for neighbor discovery and verification. The maximum hop for neighbor discovery is 1.
    \item The node multicasts to all routing nodes (except neighbor nodes) stored in the current routing node, i.e., vertices of all records on the Koch curve. The maximum hop for the remote router is $2\times t, t\neq 0$, the tier number. The transmissions of packets are designated to the router vertices to avoid flooding in the network and remove the necessity of pinging every router along the way to the destination.  
    \item The node broadcasts to all local entities' records stored in the current routing node. The maximum hop for reaching the entity is 1.
\end{enumerate}

For each successful transmission with a response received by the current routing node, the length of routes is calculated in terms of the total segment length, $s$, which is the minimum unit of SnowedNet length defined as the shortest side length. In practice, the segment length is equivalent to 1 hop between two highest-tier neighboring nodes, i.e., the shortest segment in the fractal shape.

\subsubsection{Routing strategy}
Having the segment and vertices defined as the number of records and routing nodes, the SnowedNet is illustrated in Fig. \ref{fig:hyper}. By comparing the snowflake shape with the network characteristics of SnowedNet, we can conclude the following attributes: tier number $t$; node numbers $N$; the length of segments kept by the highest tier node noted as $s$; segment length $S$; the number of blocks per segment $b$; block numbers $B$; the number of records per block $r$; and record numbers $R$.

The number of nodes $N$ can be defined with the given number of tiers $t$,
\begin{equation}
  N = 3\times 4^{t}.
\end{equation} 

The length of total segments $S$ for a given number of tiers $t$ is,
\begin{equation}
  S = 3\times((\frac{4}{3})^{t}).
\end{equation}

The total number of blocks $B$ in the above SnowedNet for one SnowedNet epoch and the total records $R$ kept in the given SnowedNet with $r$ records in each block are
\begin{equation}
  B = b \times S, \and   
  R = r \times B.
\end{equation}

%To define a SnowedNet, following parameters are required: 
%\begin{itemize}
%   \item Tier number $t$, is to determine how many tier of fractal is generated for the network, it will yield the total node number $N$.
%   \item Segment length $s$, also known as shortest side length, is to determine how long the highest tier triangle's side length. The full length of Koch snowflake can be represented by how many segments in total. Hence, it is easy to know the capacity of given network.
%   \item Block number $b$, is the number of blocks on each segment, by multiplying with total segments 
% \end{itemize}

The blockchain records contain the information needed for routing and switching, which are essential self-claimed identities from clients of their current network address bindings. In the case of switching, the local record utilizes the binding of the entity's encrypted address, i.e., $\sf{BCADD}$, and the network interface, e.g., the port number of a switch. By having the $\sf{BCADD}$ as the header, the endpoint switch/router is able to steer the traffic between any entities tagged within the switch/routing table records and the attached interfaces. 
% If the entity is not located in the same router/switch domain, the switching node automatically becomes the routing node, which routes the traffic using the SnowedNet address. The link state information can be eventually learned by the home router. Meanwhile, the router chooses the dispatch mode, which can adopt Quality of Service (QoS) options, e.g., DiffServ and IntServ. In addition to QoS, the dispatch can use duplicated or triplicated (more duplicates are possible with higher $D$ in higher dimensional network) packets to ensure higher reliability, as the router is connected to at least two or more same-tier or lower-tier routers in the advanced state. 

As a part of blockchain-integrated functionality in Web 3.0, smart contracts can be directly involved in making Service Level Agreement (SLA) deals between two endpoint users. Such SLA information is readable to all routers, and acknowledgments of resource reservation can be collected by replying to the request hosted by smart contracts.

\section{Verifiable Identities between Reality and Web 3.0}\label{sec:operate}

As aforementioned, crypto-network-entity, a network avatar of the user's credential, is native to Web 3.0, existed in the fractal polytope shape of a decentralized network, and its identifiers are used by dApp, DeFi, Decentralized Autonomous Organizations (DAOs) and decentralized infrastructures. On the other hand, identities should be consistent in both decentralized and centralized infrastructures. Therefore, there is a requirement of consistent verifiable identities between Web 3.0 and reality, satisfying the legal anonymity and the linkability of decentralized identifiers from network to applications.

\subsection{Overview}
\begin{figure}[htbp]
    \centering
    \includegraphics[width =0.5\textwidth]{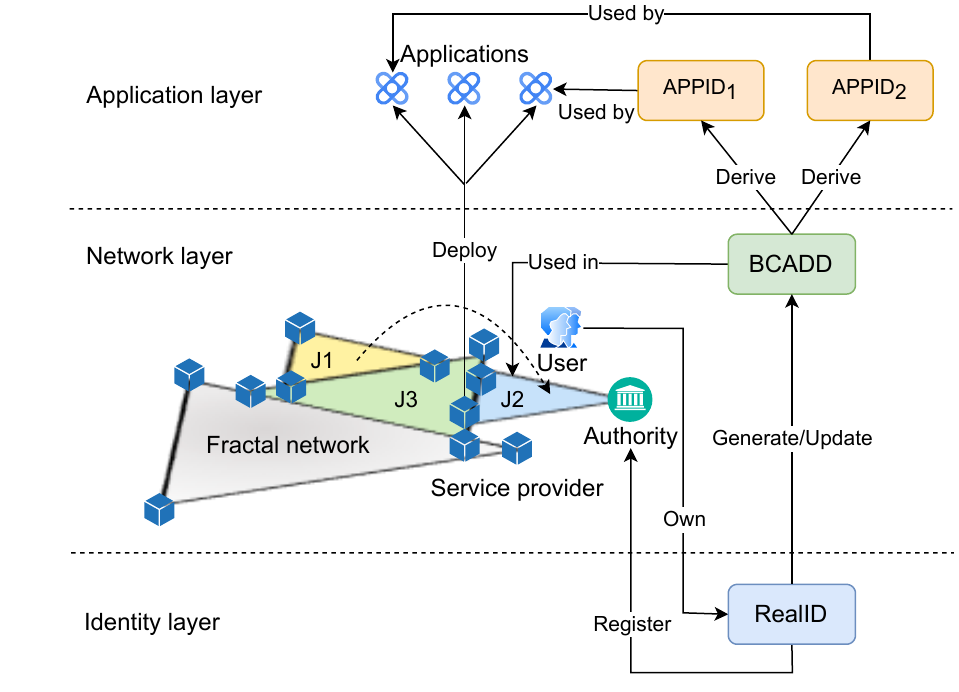} % to be replaced by vector img
    \caption{Three-tier identity scheme for the Web 3.0 fractal network.}
    \label{fig:identity}
    %\vspace{-12px}
\end{figure}

To ensure user anonymity in the Web 3.0 network, a three-tier identity mechanism was introduced in deController~\cite{Xu2023} to connect entities in reality with anonymous identifiers in the network and services for security and privacy. 
According to the identity generation procedure, the authorities can audit anonymous identities in Web 3.0 network. However, the public should have no information about the identity linked to the anonymous identifiers.%, which causes disconnection between identity in real world and digital identifiers in network. 
Moreover, once the identifiers generation procedure involves more parameters from networks or services, auditing from authority becomes a challenge. Here, we follow the design of the three-tier identity, refine the parameters in identity derivation, and enable the public verifiable authenticity of the privacy-preserving links between identities in Web 3.0 and reality as illustrated in Fig. \ref{fig:identity} and Fig. \ref{fig:workflow}.
%%%%To realize anonymous identity and equivalent identity verification in Web 3.0, an identity scheme is essential to link identities in reality to identities used in Web 3.0. Furthermore, such identities in Web 3.0 should be anonymous to protect user privacy and easy to be verified. Therefore, we propose a multi-layer identity scheme with three tiers consisting of identity in reality, Web 3.0 identity, virtual identity illustrated as follows.

\subsection{Identity and Operations}
All the identifiers should not reveal any sensitive identity information. As shown in Fig. \ref{fig:identity}, when the user moves from one jurisdiction (J1) to another (J2), the identifiers $\sf{BCADD}$ and $\sf{APPID}$ are generated from the registered $\sf{RealID}$ and privacy-preserving invisible links are built among them. $\sf{BCADD}$ and $\sf{APPID}$ can be verified by the public as seen in Fig. \ref{fig:workflow} to allow people to confirm the $\sf{BCADD}$ and $\sf{APPID}$ provided by the user are valid when the user uses them in the Web 3.0 cyberspace, e.g., network, applications, services, etc.
%%%%To implement such an identity scheme shown in Fig. \ref{fig:identity}, some operations are required in both Web 3.0 cyberspace and reality.

%Identity generation is operated by the user itself according to the parameters received from authorities and applications. The validity of anonymous identities is open to public verification (including authority). Therefore, 

%One promising technique to achieve the identity operation is zero-knowledge proof (ZKP).

\begin{itemize}
    %\item \textit{Identity in reality:} noted by $\sf{ID}$ is the identity of the user in reality including all identifying information of the user. $\sf{ID}$ is unchangeable in the Web 3.0 cyberspace and should be issued, updated, and verified by the authorities from its country of origin or other permitted authorities in reality.

    \item \textit{Identity credential:} noted by $\sf{RealID}$ is the master key of the user in the Web 3.0 identity layer implicitly linked to the identity $\sf ID$ of the entity in reality, which is unchangeable in the Web 3.0 cyberspace and should be issued, updated, and verified by the authorities from its country of origin or other permitted authorities in reality. $\sf RealID$ is the root of all the identifier derivation in the Web 3.0 cyberspace.

    \item \textit{Web 3.0 identity:} %%%%noted by $\sf{WID}$ is the identity of the user in Web 3.0 cyberspace. $\sf{WID}$ can be generated and updated by the user itself based on $\sf{ID}$ and verified in public. In addition, sensitive identity information should be protected in $\sf{WID}$.
    noted by $\sf{BCADD}$ is the identity of the user in the network layer. $\sf{BCADD}$ can be generated and updated by the user itself to be derived with optional parameters $\sf para_{auth}$ from authorities based on $\sf RealID$. The $\sf{BCADD}$ and its link to the corresponding $\sf RealID$ can be verified by the public.
    
    \item \textit{Virtual identifiers:} %%%%noted by $\sf{VID}$ represents virtual identities used in different Web 3.0 services and applications, which is linked to $\sf{WID}$ and can be verified in public. $\sf{VID}$ is generated by the user only containing essential information for the specific usage such as balance check for services and age confirmation for applications.
    noted by $\sf{APPID}$, represents users' identities in the application layer for different services and applications. Generally, $\sf APPID$ is derived from $\sf BCADD$ together with optional parameters $\sf para_{auth}'$ from authorities and the corresponding service/application. Furthermore, it can contain essential user information for specific usage such as balance checks for services and age confirmation for applications. 
\end{itemize}

%First of all, since $\sf{BCADD}$ should not reveal any sensitive identity information, it needs to be generated from $\sf{Credential}$ to establish an invisible link between $\sf{Credential}$ and $\sf{BCADD}$. Such a link is updatable to ensure new $\sf{BCADD}$ can be generated when the corresponding identity information of $\sf{Credential}$ is changed. The next operation is identity derivation to generate $\sf{VID}$ from $\sf{BCADD}$. This identity derivation is operated by users to generate and link a series of $\sf{VID}$ to $\sf{BCADD}$ depending on the required information by different services and applications. For example, if an application in Web 3.0 only needs to verify the age of the user who is trying to access, the user derives a $\sf{VID}$ from $\sf{BCADD}$ containing only the age information for the age check without offering any other redundant identity information in $\sf{VID}$ to avoid other identity information leakage.

%Moreover, $\sf{BCADD}$ and $\sf{VID}$ can be verified by others in public to allow them to confirm the $\sf{BCADD}$ and $\sf{VID}$ provided by the user are valid when the user uses them in the Web 3.0 cyberspace, e.g., network, applications, services, etc. Due to the invisibility of the identity information behind $\sf{BCADD}$ and $\sf{VID}$, sensitive identity information can be kept secret in the verification.

There are three distinct requirements that should be considered to implement the proposed three-tier identity scheme including identity anonymity, linkability between two tiers and public verification. However, conventional certificates and other public-key methods are difficult to meet all three requirements in a fully decentralized infrastructure.%Some new technical routes should be explored and introduced to implement the proposed three-tier identity scheme in the Web 3.0 scenario.

To achieve the required identity operations, we have two research questions: 
1) How to verify the users' ownership of anonymous identifiers? 2) How to verify anonymous identifiers are generated using authority-provided parameters, hence approved by authorities?

\subsection{Technical route: Non-Interactive Zero-Knowledge Proof (NIZK)}
One promising technical route could be zero-knowledge proof (ZKP),
%, which has been comprehensively investigated and advanced. ZKP stemmed from an ancient conundrum of a millionaire -- how to prove his richness without showcasing his assets. ZKP can be regarded as a circuit for one party to prove the truth of the statement to another party. The highlight of ZKP is to realize the proof of a statement without revealing anything other than the statement itself is true i.e., no leakage of the statement-related information.
first proposed by Goldwasser, Micali, and Rackoff in 1985 \cite{GMR85}. The highlight of ZKP is one party proves to another party using cryptographic methods that it possesses knowledge without revealing any information about the knowledge.

%After vigorous discussion and studies in recent years, 
There are two well-established categories of ZKP: interactive and non-interactive. The interactive ones enable the prover to show its enough knowledge of the statement to the verifier via interactive procedures. However, such a design requires both the prover and the verifier keeping online and probably incurs heavy communication overhead.

To mitigate the communication complexity of interactive ZKP, non-interactive zero-knowledge (NIZK) \cite{fiat1986prove} is proposed to prune the interactive rounds to one round, which is more desirable than interactive ZKP in the Web 3.0 scenario. A general NIZK system is sculpted by three algorithms: $(\sf Setup, Prove, Verify)$ with a statement $x$ and witness $w$.
 (1) ${\sf Setup}(1^\kappa)\rightarrow\sigma$, where $\sigma$ is a public random string; (2) ${\sf Prove}(\sigma, x, w)\rightarrow \pi$, where $\pi$ is the proof; (3) ${\sf Verify}(\sigma, x, \pi)\rightarrow 0/1$, where 1 means accepts the proof and 0 means rejects. 

To apply the NIZK scheme to our refined identity scheme, the algorithm generating ${\sf BCADD}$ from ${\sf RealID}/{\sf para_{auth}}$ or the algorithm generating ${\sf APPID}$ from ${\sf BCADD/para_{auth}'}$ is regarded as the statement $x$ in NIZK. Meanwhile, ${\sf RealID/para_{auth}}$ and ${\sf BCADD/para_{auth}'}$ are regarded as witnesses in different statements. To implement the ${\sf Setup}$ procedure, a smart contract or the authorities can be regarded as trusted parties to generate the public random string. Then, the user performs ${\sf Prove}$ and includes the proof $\pi$ into auxiliary values ${\sf aux_{BCADD}/aux_{APPID}}$. The verification procedure can be implemented as smart contracts in Web 3.0 so all $\sf{BCADD}$ and $\sf{APPID}$ can be verified by the public. 
 %%%%    \begin{itemize}
 %%%%\item $Setup(pp) \rightarrow (S_p,S_v)$: this algorithm takes the public parameters ($pp$) to generate certain parameters for \textit{Prove} ($S_p$) and \textit{Verify} ($S_v$) to use later.

 %%%%\item $Prove(S_p,x,w) \rightarrow \pi$: this algorithm is invoked by the prover to take $S_p$, a public challenge $x$ from the verifier, and the secret knowledge (witness) $w$ to generate the proof $\pi$.

 %%%%\item $Verify(S_v,x,{\pi})$: the verifier can determine to accept or reject the proof $\pi$ by invoking this algorithm to verify the proof $\pi$ with using $S_v$ and $x$.
 %%%%\end{itemize}

%Nowadays, numerous NIZK protocols and zero-knowledge Succinct Non-interactive ARguments of Knowledge (zk-SNARKs) protocols have been employed on blockchain to achieve private transactions, verifiable computations, and decentralized identity and authentication\cite{Candid}. %However, none of them touches identities operation Here, we give a brief introduction of a generic NIZK algorithm and show how well the algorithm can be applied to our identity operation.

\subsection{Three-tier identity based on NIZK}

\begin{figure}[htbp]
    \centering
    \includegraphics[width=0.5\textwidth]{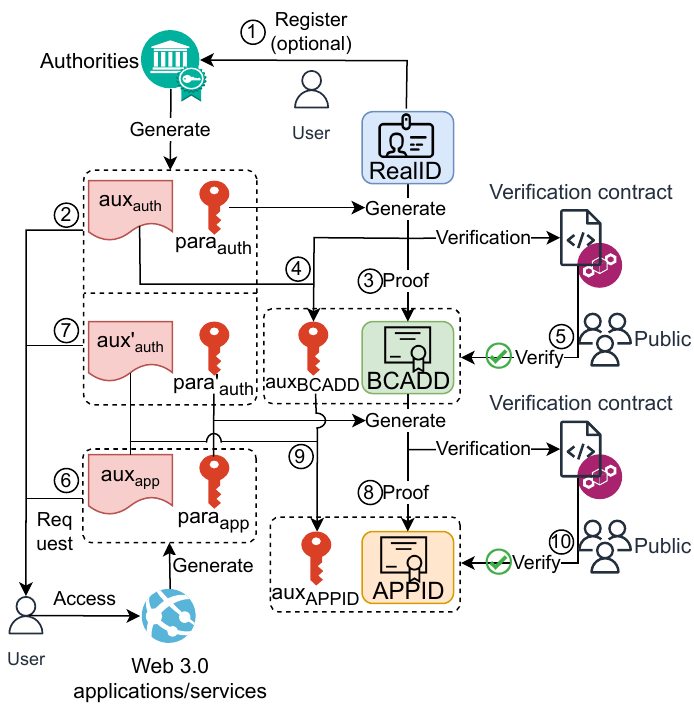}
    \caption{Workflow of the three-tier identity scheme based on NIZK.}
    \label{fig:workflow}
    %\vspace{-10px}
\end{figure}

Though $\sf BCADD$ can be derived directly from $\sf RealID$, it needs to meet the legal requirement as the authority may need to attain the user's identity. Therefore, the authority needs to first bind $\sf RealID$ to $\sf ID$ and urge the use of $\sf BCADD$ derivation with the provided parameters. We present an example of how public verifiable anonymous identities are generated, used, and verified in Web 3.0 as demonstrated in Fig. \ref{fig:workflow}.

\subsubsection{From ${\sf RealID}$ to $\sf BCADD$} First, user registers $\sf{RealID}$ to authority as shown in Fig. \ref{fig:workflow}, $\textit{step}$~1. Then, the user receives parameter $\sf para_{auth}$ with auxiliary value $\sf aux_{auth}$ from an authority, as shown in Fig. \ref{fig:workflow}, $\textit{step}$~2, where $\sf aux_{auth}$ cannot be forged by other parties. $\sf aux_{auth}$ can be regarded as signature of $\sf para_{auth}$ and $\sf{RealID}$ to endorse their authenticity. In $\textit{step}$~3, user generates $\sf BCADD$ from $\sf{RealID}$ and $\sf para_{auth}$. Then, in $\textit{step}$~4, user generates $\sf aux_{BCADD}$ from $\sf aux_{auth}$, $\sf para_{auth}$, and $\sf RealID$. Note that both $\sf BCADD$ and $\sf aux_{BCADD}$ perfectly hide $\sf{RealID}$. $\sf aux_{BCADD}$ can be used to verify the validity of the parameters in generating $\sf BCADD$. Each time user uses $\sf BCADD$ in the network, it attaches $\sf aux_{BCADD}$, as shown in $\textit{step}$~5, which enables the public to verify that $\sf BCADD$ is generated from a real $\sf RealID$ that is already registered to authority.
    
\subsubsection{From ${\sf BCADD}$ to $\sf APPID$} When a user access an application in Web 3.0 and this application needs to verify the private data held by the user, e.g., the age of the user, it sends a verifying request with $\sf para_{app}$ and auxiliary value $\sf aux_{app}$ to the user, as shown in Fig. \ref{fig:workflow}, $\textit{step}$~6. 
    Since the age information is personal information should be held by authorities, the user should receive $\sf para_{auth}'$ with $\sf aux_{auth}'$, as shown in $\textit{step}$~7, where $\sf para_{auth}'$ contains user age and other optional parameters. Then, the user derives $\sf{APPID}$ from $\sf BCADD$, $\sf para_{app}$, $\sf para_{auth}'$ in $\textit{step}$~8, and generates $\sf{aux_{APPID}}$ from $\sf aux_{BCADD}$, $\sf aux_{app}$, $\sf aux_{auth}'$, $\sf para_{app}$, $\sf para_{auth}'$ and $\sf BCADD$ in $\textit{step}$~9. Noted that both $\sf APPID$ and $\sf{aux_{APPID}}$ hides $\sf BCADD$, $\sf RealID$ and user age. If user attaches $\sf{aux_{APPID}}$ with $\sf{APPID}$, the public could verify that $\sf{APPID}$ is linked to a valid $\sf BCADD$ or $\sf ID$ by checking $\sf aux_{APPID}$ in $\textit{step}$~10. 

    In this way, we can establish a holistic identity scheme, extending $\sf RealID$ to $\sf BCADD$ in the network layer and $\sf BCADD$ to $\sf APPID$ in the application layer so a joint authentication can be achieved by the network for applications via decentralization-ready physical infrastructures, e.g., blockchain network routers and mobile networks \cite{Xu2023decent}.%, as long as the identity is verifiable by the service providers via decentralization-ready physical infrastructures, e.g., blockchain network routers and mobile networks \cite{Xu2023decent}.

\section{Challenges and Opportunities}
\subsection{Questioning the existence of crypto-network-entity}
The definition of new ``existence'' brings rigorous discussion on the juridical aspect. In the discovered network, the existence comes after the crypto-loaded physical entity is plugged into the network, hence leaving the record of crypto-network-entity on the blockchain ledger. Therefore, the challenges to new world legal governance become an extension of the cyber or physical world, depending on where in the cyber or physical world the crypto-network-entity is plugged in. The physical location of vertices nodes in the fractal simplex network is also subject to the local jurisdiction, and subjects to the real physical world entity. In this way, people may wonder whether the network is independent or not, and the answer can be straightforward by examining the existence when physical entities are removed. The status of the network is independent from any single networking device but subjects to the decentralized infrastructure as a whole. However, the status of the network does not undermine the existence of crypto-network-entity as long as the decentralized information medium stays true.  

\subsection{Questioning on  network calculus and pathfinding}
In the fractal regular polytope network, we have the shortest path known as $2t$, but the stub of the network has entered an extremely complex space where routes need to be found in hyperdimension.  
More advanced manifold-based geometric solutions to the shortest route on the fractal polytope network are critical challenges and future work for the comprehensive simplex fractal network.
Furthermore, more basic properties such as network mass, density, volume, acceleration, power, etc., can be mapped into the fractal polytope network for a better understanding of network characteristics.   

\subsection{Questioning the privacy and legal compliance of anonymity}
Due to the nature of user anonymity in Web 3.0, it does not set any restriction to the real identity behind the user which may result in abuse and forgery of user identities. Therefore, the supervision of $\sf{RealID}$ is a key point to be considered to align the proposed three-tier identity scheme to legislation in building the global Web 3.0 cyberspace. As shown in Fig. \ref{fig:workflow}, an optional design is to utilize authorities to issue and update a user's $\sf{ID}$ to avoid the user forging its identity information or using identity information of others without permission. When users try to access certain Web 3.0 resources, in which real-name registration is required, such a design can ensure the generated $\sf{BCADD}$ from $\sf{RealID}$ and the derived $\sf{APPID}$ from $\sf{BCADD}$ contain authorized identity information to represent the real identity of the user. 

\section{Conclusion and future work}
In this paper, we propose SnowedNet, reforming the spatial information of everything that is connected to Web 3.0 in a hyperdimensional simplex topology, timestamp everything with blockchain consensus, and extending the Web 3.0 existence to cyberspace and physical space with ZKP on their identities derivatives. The fractal-driven topology is illustrated with a 2-simplex, a regular triangle with Koch iterations, showing not only the juridical broader mapped into shards but also enormous capacity and simplicity in building the future generation decentralized network.  

% This paper has presented a novel architecture of fractal simplex network in any dimension. 
However, there are limited findings on the shortest path on higher dimensional fractal topology. In future work, a comprehensive study on hyperdimensional routing protocol with state-of-the-art polytope theories is plan to help find the shortest path in any dimension of fractal simplices. Furthermore, NIZK verifiable identities and identifier derivation need to be explored to enhance the efficiency of identifier generation to crystallize a comprehensive cryptographic scheme. In addition, NIZK identity statement should be compatible with various structures of identity information from different jurisdictional authorities.
% In this paper, we propose deController, a perspective of Web3.0 architecture for future decentralized Web3.0 infrastructures, consisting of overlay and underlay to catalyze more free and fair web access for people. The functions of deController are illustrated in a top-down sculpture of Web3.0 architecture with the considerations of concealed identity, security and privacy, and law. The term Web3.0 shall also enable not only the decentralization of giant Internet companies, but also the decentralization from the de-facto centralized infrastructure controller. 
% Our solution proposed in this paradigm can be a potential starting point for the real Web3.0 infrastructure investment, which allows the true ownership of Web3.0 beyond the content. 

\bibliographystyle{IEEEtran}
\bibliography{references}
\end{document}